\documentclass[aps,prl,reprint,twocolumn,amsmath,amssymb]{revtex4-1}
\usepackage{stmaryrd}
\usepackage{times}
\usepackage{graphicx}
\usepackage{dcolumn}
\usepackage{bm}
\begin{document}
\title{Giant Magneto-Seebeck Effect in Spin Valves}
\author{X. M. Zhang}
\affiliation{Beijing National Laboratory for Condensed Matter Physics, Institute of Physics,
Chinese Academy of Sciences, Beijing 100190, China}
\author{C. H. Wan}
\email[Electronic mail: ]{wancaihua@iphy.ac.cn}
\affiliation{Beijing National Laboratory for Condensed Matter Physics, Institute of Physics,
Chinese Academy of Sciences, Beijing 100190, China}
\author{Z. H. Yuan}
\affiliation{Beijing National Laboratory for Condensed Matter Physics, Institute of Physics,
Chinese Academy of Sciences, Beijing 100190, China}
\author{H. Wu}
\affiliation{Beijing National Laboratory for Condensed Matter Physics, Institute of Physics,
Chinese Academy of Sciences, Beijing 100190, China}
\author{Q. T. Zhang}
\affiliation{Beijing National Laboratory for Condensed Matter Physics, Institute of Physics,
Chinese Academy of Sciences, Beijing 100190, China}
\author{X. Zhang}
\affiliation{Beijing National Laboratory for Condensed Matter Physics, Institute of Physics,
Chinese Academy of Sciences, Beijing 100190, China}
\author{B. S. Tao}
\affiliation{Beijing National Laboratory for Condensed Matter Physics, Institute of Physics,
Chinese Academy of Sciences, Beijing 100190, China}
\author{C. Fang}
\affiliation{Beijing National Laboratory for Condensed Matter Physics, Institute of Physics,
Chinese Academy of Sciences, Beijing 100190, China}
\author{X. F. Han}
\email[Electronic mail: ]{xfhan@iphy.ac.cn}
\affiliation{Beijing National Laboratory for Condensed Matter Physics, Institute of Physics,
Chinese Academy of Sciences, Beijing 100190, China}

\date{\today}
\begin{abstract}
Giant magneto-Seebeck (GMS) effect was observed in Co/Cu/Co and NiFe/Cu/Co spin valves. Their Seebeck coefficients in parallel state was larger than that in antiparallel state, and GMS ratio defined as ($S\mathrm{_{AP}}$-$S\mathrm{_{P}}$)/$S\mathrm{_{P}}$ could reach -9\% in our case. The GMS originated not only from trivial giant magnetoresistance but also from spin current generated due to spin-polarized thermoelectric conductivity in ferromagnetic materials and subsequent modulation of the spin current by different spin configurations in spin valves. Simple Mott two-channel model reproduced a -11\% GMS for the Co/Cu/Co spin valves, qualitatively consistent with our observations. The GMS effect could be applied simultaneously sensing temperature gradient and magnetic field and also be possibly applied to determine spin polarization of thermoelectric conductivity and Seebeck coefficient in ferromagnetic thin films.
\end{abstract}
\maketitle
Spin caloritronics, focusing on interplay between thermoelectric transport and spin configurations of ferromagnetic materials or related hybrid structures, has extensively reshaped landscape of recent spintronics research, since discovery of spin Seebeck effect\cite{Uchida,Bauer,Kikkawa,Qu,Huang,Wu}, spin-dependent Seebeck effect\cite{Bauer,Dejene,Slachter} and spin-dependent tunneling Seebeck effect\cite{Walter,Boehnke,Lin}. The former two effects have shed light on effective methods of generating a steady pure spin current into nonmagnetic materials through wasted heat which is being abundantly produced and wasted in current microelectronics. Thereafter marriage of thermally generated spin current and varied spin configurations in magnetic tunnel junctions has bred spin-dependent tunneling Seebeck effect. However, another version of Seebeck effect in broadly applied spintronic structure, namely, giant magnetoresistance (GMR) spin valves, which deals with spin-configuration-dependent thermoelectric transport in diffusive manner, has still not been systematically researched. Gravier\cite{Gravier} and Sakurai\cite{Sakurai} $et$ $al$. reported a field-dependent thermopower in Fe/Cr and Co/Cu multilayers and they discussed little on physics hidden behind it. Very recently, Jain\cite{Jain} $et$ $al$. reported a tiny field-dependent Seebeck effect in CoFe/Cu/CoFe/IrMn/Ru spin valves. They further attributed its field dependence to the trivial origin that different resistances in parallel (P) state and antiparallel (AP) state resulted in different Seebeck coefficients accordingly. Furthermore, Hu\cite{Hu} $et$ $al$. reported a small field-dependent Seebeck coefficient in nonlocal lateral spin valve which showed even different sign of ($S\mathrm{_{AP}}$-$S\mathrm{_{P}}$)/$S\mathrm{_{P}}$ with other literatures [12, 13]. Magneto-Seebeck ratios [$\mathrm{GMS}$$\equiv$($S\mathrm{_{AP}}$-$S\mathrm{_{P}}$)/$S\mathrm{_{P}}$] reported in these researches not only diverged in sign but also in absolute values ranging from 0.1\% to 10\%. In this letter, we will demonstrate that Seebeck effect of spin valves not only exhibit specific field dependence as that of GMR effect but also its GMS ratio could be giant and even reach a higher value than GMR ratio. Furthermore, the GMS ratio has different sign with GMR in our case, which could no longer be explained as a parasitic effect of the GMR effect. We will also show physical origin behind this phenomenon in this letter afterwards and give a probable explanation on sign change in different material systems.

As shown in Fig. 1(a), spin valves with structure of SiO$_{2}$//Ni$_{81}$Fe$_{19}$(3)/Cu($t$)/Co(4)/Ir$_{22}$Mn$_{78}$(12)/Pt(3) (nominal thickness in nanometer) were grown by ULVAC magnetron sputter system with a base pressure of 1.0$\times$10$^{-6}$ Pa at room temperature. During deposition, a field of 200 Oe was used to induce an easy axis (EA) as well as exchange bias (EB) along $x$ in the spin valves. Different samples with $t$=2, 4, 6, 8 correspond to Sample 1-4. All films were then patterned into 20 $\mu m\times$270 $\mu m$ bars as shown as the dark green bar in Figure 1(b) by ultraviolet lithography and subsequent argon ion milling. Then a bar made of Cu(20)/Au(20) between Pad C and Pads D as well as Pad A was deposited to transport heating current and detect Seebeck voltages, respectively. As measuring magneto-Seebeck effect, we applied a heating current $I$=$I$$_{0}$sin($\omega t$) by Keithley 6221 between Pad C and Pad D along $y$ which produced a thermal gradient along $x$ ($\nabla_x$$T$). In this direction, no net current flowed between Pad A and Pad B since the bar made of spin valve (SV) were isolated from the heating bar by 7 $\mu$m. Pad A and Pad B were just used to measure Seebeck voltages along the SV. Seebeck voltage was first amplified by a low-noise preamplifier (SR560, Standford) with gain of 20000. Then SR830 only measured locked-in voltage signals $V\mathrm{_{S}}$=$V\mathrm{_{S0}}$cos(2$\omega t$) to improve signal-to-noise ratio and also to rule out voltage origins such as Peltier effect and other higher order signals. Magnetic field was provided by Physical Properties Measurement System (PPMS-9T, Quantum Design). Environment temperature during measurement was kept constantly as 300 K by PPMS. We have verified that GMS was indeed independent on $\omega$ though the magnitude of $V\mathrm{_{S0}}$ did (SI A). In the following, we will only present results measured at $\omega$=731.7 Hz. $M$-$H$ hysteresis loop was measured by vibrating sample magnetometer (EV9, MicroSense).
\begin{figure}
\includegraphics[width=0.45\textwidth]{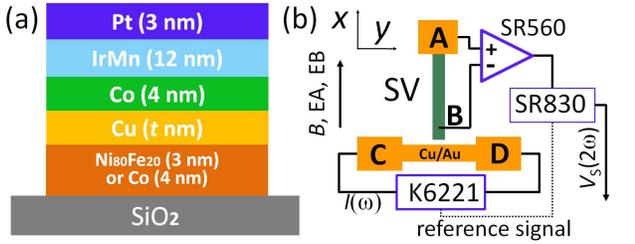}%
\caption{\label{Fig.1} (Color online) (a) structures of spin valves and (b) pattern of devices and second harmonic measurement setups.}
\end{figure}

Fig. 2(b) shows mainly the dependence of $V\mathrm{_{S0}}$ on heating current $I_{0}$ for Sample 3 ($t$=6). Fig. 2(a) shows the field dependence of two-terminal resistance between Pad A and Pad B as applied $B$ along $x$. Exchange bias of the Co pinned layer could be clearly resolved with a bias field of about 200 Oe. The hysteresis loop around zero fields originated from the NiFe free layer. Fig. 2(a) also shows a natural result that resistance of Sample 3 in antiparallel state (AP) was larger than that in parallel state (P), as normal GMR sandwiches. Interestingly, very similar to the field dependence of resistance, a field-dependent Seebeck voltage was also observed. However, Seebeck voltage in AP state was remarkably smaller than that in P state. Here we have defined $GMR$($B$)$\equiv$[$\rho$($B$)-$\rho$(0)]/$\rho$(500 Oe) and $GMS$($B$)$\equiv$[$S$($B$)-$S$(0)]/$S$(500 Oe)=[$V\mathrm{_{S0}}$($B$,$I_{0}$)-$V\mathrm{_{S0}}$(0,$I_{0}$)]/$V\mathrm{_{S0}}$(500 Oe,$I_{0}$) with $\rho$($B$) and $S$($B$) being resistivity and Seebeck coefficient at $B$. For Sample 3, $GMR$(500 Oe)=1.5\% while $GMS$(500 Oe)=-9\%. The value of GMS was not only much higher than that of GMR, but also much higher than other thermoelectric signals such as anisotropy Seebeck effect as shown in the following results. Therefore this magneto-Seebeck effect was named as giant magneto-Seebeck effect. Fig. 2(c) shows that Seebeck voltage in P state linearly depended on heating power ($P$) as expected while the value of $GMS$(500 Oe) was nearly independent on the $P$. It was also significant that exchange bias field of the Co layer was gradually reduced due to the increase in average temperature of the sample. Higher temperature usually results in higher switching probability in the same field due to higher thermal perturbation as reported in Ref. \cite{Stiles, Gloanec}.
\begin{figure}
\includegraphics[width=0.45\textwidth]{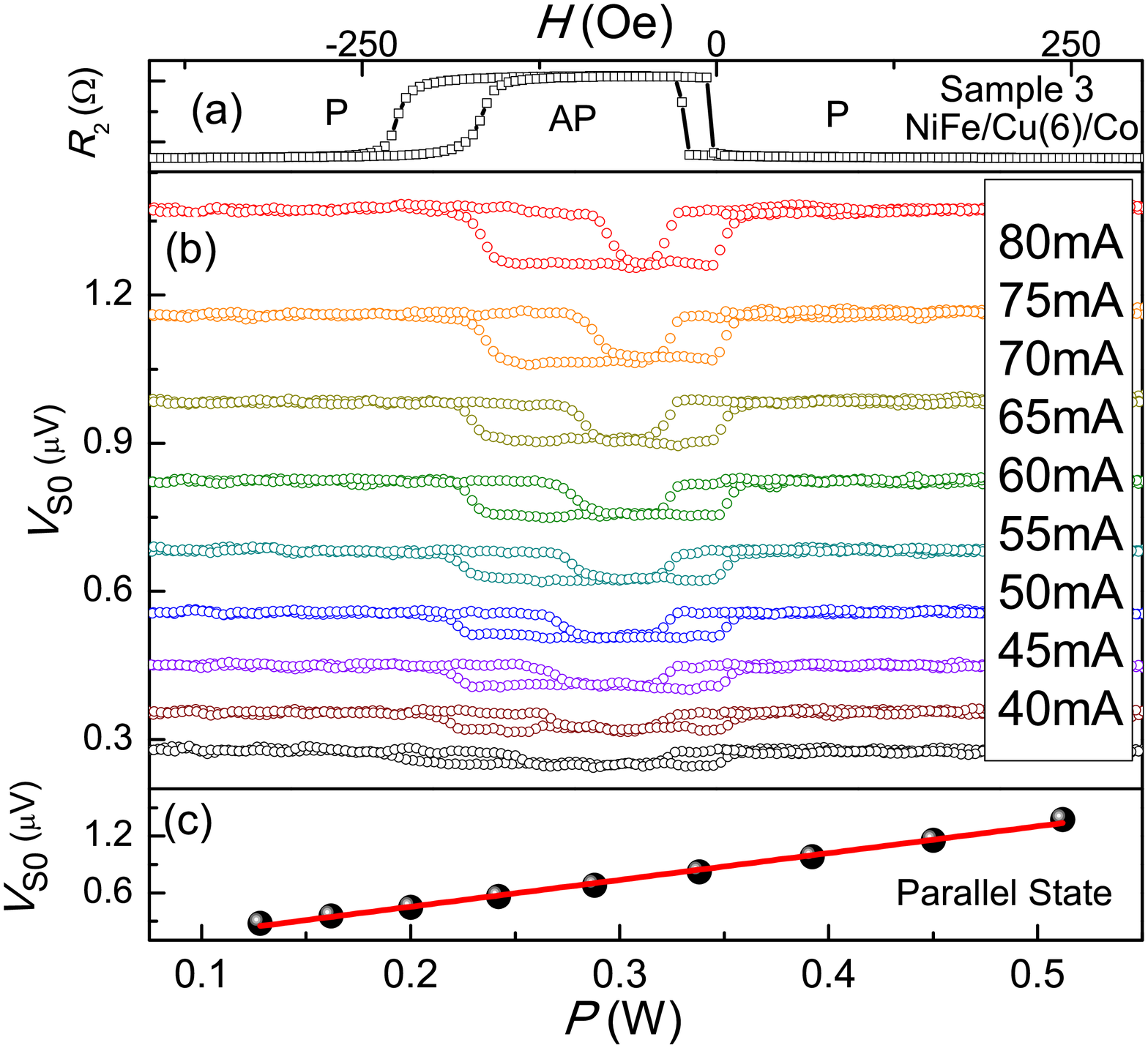}%
\caption{\label{Fig.2} (Color online) (a) two-terminal resistance between Pad A and Pad B in Sample 3, (b) field-dependent Seebeck voltages measured at elevated heating current $I_{0}$ and (c) the dependence of Seebeck voltages in parallel state on heating power $P$=$I_{0}^{2}R\mathrm{_{h}}$ where $R\mathrm{_{h}}$=80 was resistance of heating bar.}
\end{figure}

In order to rule out other possibilities of the field-dependent voltage contribution from anisotropy Seebeck effect under $\nabla_{x}T$ or anomalous Nernst effect under $\nabla_{z}T$ or planar Nernst effect under $\nabla_{y}T$, we have performed some controlled experiments. Fig. 3(a) shows $M$-$H$ hysteresis loop of Sample 1($t$=2) and Sample 2($t$=4) with $B$ along the EA and EB direction. Different from Sample 2, pinned layer and free layer in Sample 1 coupled with each others. Therefore only P state in Sample 1 was observed, which agreed with magnetotransport measurement. We only observed a tiny anisotropy magnetoresistance as well as a planar Hall voltage in Sample 1 as varying $B$ along $x$, without any signals from GMR effect (Fig. 3(b) and (c)). Meanwhile, as we applied $I_{0}$=80 mA along the heating bar and $B$ along the hard axis ($y$), only a negligible small field-dependent voltage was observed around zero fields (Fig. 3(b) and (c)). The magnitude of the field-dependent voltage was only about 15 nV, close to the noise level of 10 nV but much smaller than that from GMS effect. The latter reached over 100 nV as $I_{0}$=80 mA in Sample 3.  This tiny field-dependent Seebeck effect in Sample 1 was induced by anisotropy magneto-Seebeck effect. We also measured field-dependence of Seebeck voltages of Co(20)/Pt(3) and Ni$_{81}$Fe$_{19}$(20)/Pt(3) and also found no observable field-dependent voltages (SI B). These controlled measurements compelled us to conclude that the field-dependent voltages in Fig. 2 and Fig. 4 resulted from different Seebeck voltages between AP state and P state, instead of from planar Nernst effect or anisotropy magneto-Seebeck or anomalous Nernst effect.
\begin{figure}
\includegraphics[width=0.45\textwidth]{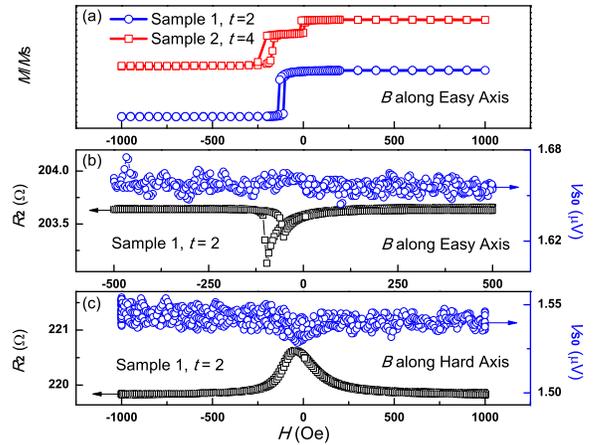}%
\caption{\label{Fig.3} (Color online) (a) $M$-$H$ curves of Sample 1 and Sample 2. The latter shows separated switching of free layer and pinned layer in Sample 2 ($t$=4) while the former shows parallelly coupled free layer and pinned layer in Sample 1 ($t$=2). Anisotropic magnetoresistance (black and empty squares) and thermoelectric voltages (blue and empty circles) of Sample 1 as (b) $B$ along easy axis and (c) $B$ along hard axis.}
\end{figure}
GMS and GMR effects were also detected in another spin valve SiO$_{2}$//Co(4)/Cu(4)/Co(4)/Ir$_{22}$Mn$_{78}$(12)/Pt(3) (Fig. 4). GMR and GMS ratios of the Co/Cu/Co spin valve were, respectively, 4.6\% and -6.0\%, both higher than 2.6\% and -4.5\% of Sample 2. Though different in values, these spin valves shared some characteristics in magneto-electrotransport properties and magneto-thermotransport properties. Firstly, their GMR ratios were all positive as usual while their GMS ratios were always negative. Results measured at low temperatures also showed the same trends (SI C). This universal phenomenon was also observed in Sample 4 (not shown here). The negative GMS ratios meant that $|$$S\mathrm{_{AP}}|<|S\mathrm{_{P}}|$. Secondly, their GMR ratios were all smaller than their GMS ratios in absolute values. In the following discussion, we will introduce physical origins beneath this universality based on a simple linear response theory in which temperature gradient and electric field independently influence charge and spin transport inside SV and in which they do not affect transport coefficients such as conductivity and thermoelectric conductivity of the system.
\begin{figure}
\includegraphics[width=0.45\textwidth]{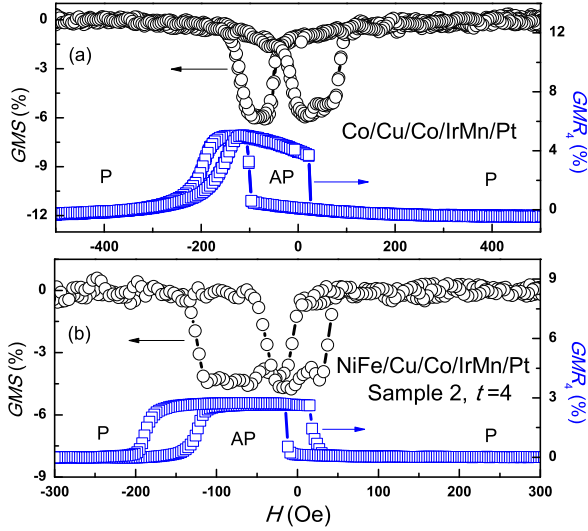}%
\caption{\label{Fig.4} (Color online) (field dependence of GMS and GMR4 (measured in four-terminal geometry) in (a) Co/Cu/Co/IrMn/Pt and (b) NiFe/Cu/Co/IrMn/Pt spin valves}
\end{figure}
Mott two-spin-channel model\cite{Mott,Fert:1190,Fert:184420} which assumes much longer spin relaxation length than mean-free path in both ferromagnetic and nonmagnetic layers has been broadly applied in estimating GMR values. Here we generalized the model into thermotransport region and tried to evaluate the value of GMS. For spin relaxation length in copper is several hundreds of nanometers\cite{Kimura,Jedema,Sampaio}, much higher than thickness ($t$) of copper in our spin valves, we further supposed there was nearly no spin relaxation event occurring in the copper. We also suppose the same thickness of the two FM layers as occurred in our Co/Cu/Co spin valves. Besides, we theoretically dealt with the case of current-perpendicular-plane (CPP) geometry instead of current-in-plane (CIP) geometry to grasp main physics though practical measurement was implemented in current- and temperature-gradient-in-plane geometry. In P state, spin-polarized current in the first FM layer would equal that in the second FM layer for each spin channel since few spins flip inside Cu as assumed. $J_{1}^{\uparrow/\downarrow}$=$\alpha_{1}^{\uparrow/\downarrow}\nabla T_{1}^{\uparrow/\downarrow}$=$\alpha_{2}^{\uparrow/\downarrow}\nabla T_{2}^{\uparrow/\downarrow}$=$J_{2}^{\uparrow/\downarrow}$ to fulfill continuity condition of current density in both spin channels. Here $J_{i}^{\uparrow/\downarrow}$, $\alpha_{i}^{\uparrow/\downarrow}$ and $\nabla T_{i}^{\uparrow/\downarrow}$ were spin-polarized current, thermoelectric conductivity and temperature-gradient in $i$th FM layer for $\uparrow$/$\downarrow$ spin channels, respectively. Especially, the concept of different temperatures in different spin channels has also been supposed and even experimentally testified in Ref.\cite{Dejene:839}. Total current $J$=$J_{1}^{\uparrow}$+$J_{1}^{\downarrow}$=$J_{2}^{\uparrow}$+$J_{2}^{\downarrow}$ and average temperature gradient 2$\nabla T$=$\nabla T_{1}^{\uparrow/\downarrow}$+$\nabla T_{2}^{\uparrow/\downarrow}$, neglecting temperature gradient dropped on copper. Thus
$J=[\alpha_{1}^{\uparrow}\alpha_{2}^{\uparrow}/(\alpha_{1}^{\uparrow}+\alpha_{2}^{\uparrow})+\alpha_{1}^{\downarrow}\alpha_{2}^{\downarrow}/(\alpha_{1}^{\downarrow}+\alpha_{2}^{\downarrow})]2\nabla T=2\alpha\mathrm{_{P}}\nabla T$. The $\alpha\mathrm{_{P}}$ is thermoelectric conductivity of spin valves in P state. In this case,
$\rho\mathrm{_{P}}$=($\rho_{1}^{\uparrow}$+$\rho_{2}^{\uparrow}$)($\rho_{1}^{\downarrow}$+$\rho_{2}^{\downarrow}$)/($\rho_{1}^{\uparrow}$+$\rho_{2}^{\uparrow}$+$\rho_{1}^{\downarrow}$+$\rho_{2}^{\downarrow}$).
Therefore $S\mathrm{_{P}}\equiv E\mathrm{_{P}}/\nabla T=\rho\mathrm{_{P}}J/\nabla T=2\rho\mathrm{_{P}}\alpha\mathrm{_{P}}$. In AP state, $J_{1}^{\uparrow/\downarrow}=\alpha_{1}^{\uparrow/\downarrow}\nabla T_{1}^{\uparrow/\downarrow}=\alpha_{2}^{\downarrow/\uparrow}\nabla T_{2}^{\downarrow/\uparrow}=J_{2}^{\downarrow/\uparrow}$ also to satisfy current density continuity requirement. $J=J_{1}^{\uparrow}+J_{1}^{\downarrow}=J_{2}^{\uparrow}+J_{2}^{\downarrow}$ and 2$\nabla T$=$\nabla T_{1}^{\uparrow/\downarrow}$+$\nabla T_{2}^{\downarrow/\uparrow}$. $J$=[$\alpha_{1}^{\uparrow}\alpha_{2}^{\downarrow}$/($\alpha_{1}^{\uparrow}$+$\alpha_{2}^{\downarrow}$)+$\alpha_{1}^{\downarrow}\alpha_{2}^{\uparrow}$/($\alpha_{1}^{\downarrow}$+$\alpha_{2}^{\uparrow}$)]2$\nabla T$=2$\alpha\mathrm{_{AP}}\nabla T$. Here, correspondingly, the $\alpha\mathrm{_{AP}}$ is thermoelectric conductivity in AP state. And
$\rho\mathrm{_{AP}}$=($\rho_{1}^{\uparrow}$+$\rho_{2}^{\downarrow}$)($\rho_{1}^{\downarrow}$+$\rho_{2}^{\uparrow}$)/($\rho_{1}^{\uparrow}$+$\rho_{2}^{\uparrow}$+$\rho_{1}^{\downarrow}$+$\rho_{2}^{\downarrow}$).
Thus $S\mathrm{_{AP}}$$\equiv$$E\mathrm{_{AP}}$/$\nabla T$=$\rho\mathrm{_{AP}}J$/$\nabla T$=2$\rho\mathrm{_{AP}}\alpha\mathrm{_{AP}}$. Ratio of GMS could be expressed via Equation \ref{eq.GMS1}.
\begin{equation}\label{eq.GMS1}
GMS+1=\frac{(\rho_{1}^{\uparrow}+\rho_{2}^{\downarrow})(\rho_{1}^{\downarrow}+\rho_{2}^{\uparrow})(\alpha_{1}^{\uparrow}+\alpha_{2}^{\uparrow})(\alpha_{1}^{\downarrow}+\alpha_{2}^{\downarrow})}{(\rho_{1}^{\uparrow}+\rho_{2}^{\uparrow})(\rho_{1}^{\downarrow}+\rho_{2}^{\downarrow})(\alpha_{1}^{\uparrow}+\alpha_{2}^{\downarrow})(\alpha_{1}^{\downarrow}+\alpha_{2}^{\uparrow})}
\end{equation}
It shows that GMS ratio is dependent both on spin polarization of resistivity and thermoelectric conductivity in two FM layers. For the Co(4)/Cu(4)/Co(4) spin valve, GMR and GMS could both be simplified as Equation \ref{eq.GMS2} and Equation \ref{eq.GMS3}. Equation \ref{eq.GMS2} is also derived in Ref. \cite{Edwards} and it is still applicable for CIP geometry. We thus expect Equation \ref{eq.GMS3} is also instructive enough to explain our thermotransport data in CIP geometry.
\begin{equation}\label{eq.GMS2}
GMR=\frac{P_{\sigma}^{2}}{1-P_{\sigma}^{2}}
\end{equation}
\begin{equation}\label{eq.GMS3}
GMS=\frac{P_{\sigma}^{2}-P_{\alpha}^{2}}{1-P_{\sigma}^{2}}
\end{equation}
Here $P_{\sigma}$$\equiv$($\sigma_{\uparrow}$-$\sigma_{\downarrow}$)/($\sigma_{\uparrow}$+$\sigma_{\downarrow}$) and $P_{\alpha}$$\equiv$($\alpha_{\uparrow}$-$\alpha_{\downarrow}$)/($\alpha_{\uparrow}$+$\alpha_{\downarrow}$), $\sigma_{\uparrow/\downarrow}$ being conductivity of spin-up/down channel of FM layer. $\alpha_{\uparrow/\downarrow}$=$\sigma_{\uparrow/\downarrow}S_{\uparrow/\downarrow}$. Thus $P_{\alpha}$=($P\mathrm{_{S}}$+$P_{\sigma}$)/(1+$P\mathrm{_{S}}P_{\sigma}$) and $P\mathrm{_{S}}$$\equiv$($S_{\uparrow}$-$S_{\downarrow}$)/($S_{\uparrow}$+$S_{\downarrow}$) where $S_{\uparrow/\downarrow}$ is Seebeck coefficient for spin-up/down electrons in FM. The above derivation could also be interpreted in terms of a generalized equivalent circuit as shown in Fig. 5. $V$ and $\nabla T$ independently determine drifting and diffusion current density inside SV, respectively. To the role of $\rho$ in ordinary electrotransport theory, reciprocal of thermoelectric conductivity ($\rho/S$) plays a similar role in thermoelectric transport theory. The equivalent circuit in Fig. 5 could reproduce the results of Equation \ref{eq.GMS1}-\ref{eq.GMS3}.
\begin{figure}
\includegraphics[width=0.45\textwidth]{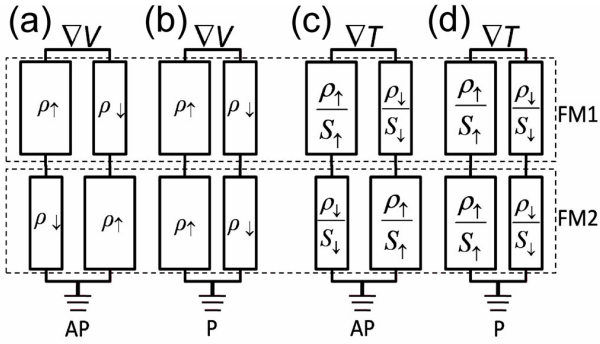}%
\caption{\label{Fig.5} (Color online) equivalent circuits driven by $\nabla V$ in (a) AP state and (b) P state and also driven by $\nabla T$ in (c) AP state and (d) P state, respectively.}
\end{figure}

Equation \ref{eq.GMS2} and \ref{eq.GMS3} shows that always $GMR\geq$0 while $GMS<$0 as $|P_{\sigma}|<|P_{\alpha}|$. For Co, $P_{\sigma}=0.4$ \cite{Sampaio} while $P_{\alpha}\simeq0.5$ \cite{Fang,Jedema}, thus ideal $GMS\simeq$-11\%$<$0 according to Equation \ref{eq.GMS3}, semi-quantitatively consistent with our experiments. $|S\mathrm{_{P}}|$ is also larger than $|S\mathrm{_{AP}}|$ in the research of Walter \cite{Walter}. If $P\mathrm{_{S}}$=0, $P_{\alpha}$=$P_{\sigma}$. Thus $GMS$=0. The nonzero GMS effect in our research indicated that Seebeck coefficient in cobalt was indeed spin-dependent as reported in Ref \cite{Fang,Jedema}. If $P_{\alpha}$=0, $GMR$=$GMS$. The observation that GMR and GMS had different signs in our spin valves also indicated that $\alpha$ in Co was spin-polarized. If $P_{\sigma}$=0, $GMR$=0 while $GMS$=-$P_{\alpha}^{2}$, meaning GMS effect could still survive even without GMR. This specialty is very different from that derived in Ref. \cite{Gravier}. The above derivation shows it was (1) different thermoelectric conductivities in different spin channels and (2) continuity condition of current density in both spin channels and also in both AP and P state that led to a giant magneto-Seebeck effect in our spin valves.

In summary, giant magneto-Seebeck and giant magnetoresistance effect were unambiguously observed in Co/Cu/Co and NiFe/Cu/Co spin valves. The former effect led to a negative GMS value which could reach -9\% in experiment and be estimated as about -11\% in a simplified two-spin-channel model. The GMS value was much higher than anisotropy Seebeck effect and other possible parasitic signals. This GMS effect was attributed to nonzero spin polarization of Seebeck coefficient of FM layers and modulation of spin-polarized current in different spin configurations. This research not only hinted a prospective application of spin valves in high sensitive temperature and magnetic field sensors but also in $T$ driven field-sensors. GMS also provided a probable manner to estimate spin polarization of Seebeck coefficient of a certain FM material which was very hard to measure in current stage.

This research has been supported by the MOST National Key Scientific Instrument and Equipment Development Projects [No. 2011YQ120053], National Natural Science Foundation of China [NSFC, Grant No. 11434014 and 51229101], Natural Science Foundation for the Youth (Grant No. 11404382), and Postdoctoral Science Foundation of China (Grant No. 2013M540154).

\end{document}


\title{Supplementary Information for Giant Magneto-Seebeck Effect in Spin Valves}
\author{X. M. Zhang}
\affiliation{Beijing National Laboratory for Condensed Matter Physics, Institute of Physics,
Chinese Academy of Sciences, Beijing 100190, China}
\author{C. H. Wan}
\email[Electronic mail: ]{wancaihua@iphy.ac.cn}
\affiliation{Beijing National Laboratory for Condensed Matter Physics, Institute of Physics,
Chinese Academy of Sciences, Beijing 100190, China}
\author{Z. H. Yuan}
\affiliation{Beijing National Laboratory for Condensed Matter Physics, Institute of Physics,
Chinese Academy of Sciences, Beijing 100190, China}
\author{H. Wu}
\affiliation{Beijing National Laboratory for Condensed Matter Physics, Institute of Physics,
Chinese Academy of Sciences, Beijing 100190, China}
\author{Q. T. Zhang}
\affiliation{Beijing National Laboratory for Condensed Matter Physics, Institute of Physics,
Chinese Academy of Sciences, Beijing 100190, China}
\author{X. Zhang}
\affiliation{Beijing National Laboratory for Condensed Matter Physics, Institute of Physics,
Chinese Academy of Sciences, Beijing 100190, China}
\author{B. S. Tao}
\affiliation{Beijing National Laboratory for Condensed Matter Physics, Institute of Physics,
Chinese Academy of Sciences, Beijing 100190, China}
\author{C. Fang}
\affiliation{Beijing National Laboratory for Condensed Matter Physics, Institute of Physics,
Chinese Academy of Sciences, Beijing 100190, China}
\author{X. F. Han}
\email[Electronic mail: ]{xfhan@iphy.ac.cn}
\affiliation{Beijing National Laboratory for Condensed Matter Physics, Institute of Physics,
Chinese Academy of Sciences, Beijing 100190, China}

\date{\today}
\maketitle

\textbf{A. Frequency dependence of Seebeck effect of Co/Cu/Co/IrMn/Pt spin valves}

We have measured frequency dependence of Seebeck voltages and GMS ratios. As shown in Figure S1, except results measured in 3.1 Hz, the Seebeck voltage in parallel state were gradually decreased with increasing frequency in our AC measurement, which might be probably induced by a lower heating efficiency in higher frequencies of heating current. However, GMS ratios were insensitive to the frequencies as shown in Figure S2, hinting this ratio was an intrinsic property of our structure or materials. No relevance between the GMS ratio and the frequency as expected also indicated effectiveness of our measurement geometry and 2nd harmonic technique.

\textbf{B. Measurement of Seebeck effect of thick NiFe and Co films}

To further confirm that giant magneto-Seebeck (GMS) effect was not due to anisotropic magneto-Seebeck effect of NiFe and Co films, we also measured Seebeck effect in NiFe and Co thick films. They were both independent on field within our measurement accuracy as shown in Figure S3. Besides, the measured ratio of $S\mathrm{_{NiFe}}/S\mathrm{_{Co}}$ was close to 1, also close to those literature values of bulk NiFe and Co \cite {Dejene}.

\begin{figure}
\includegraphics[width=0.45\textwidth]{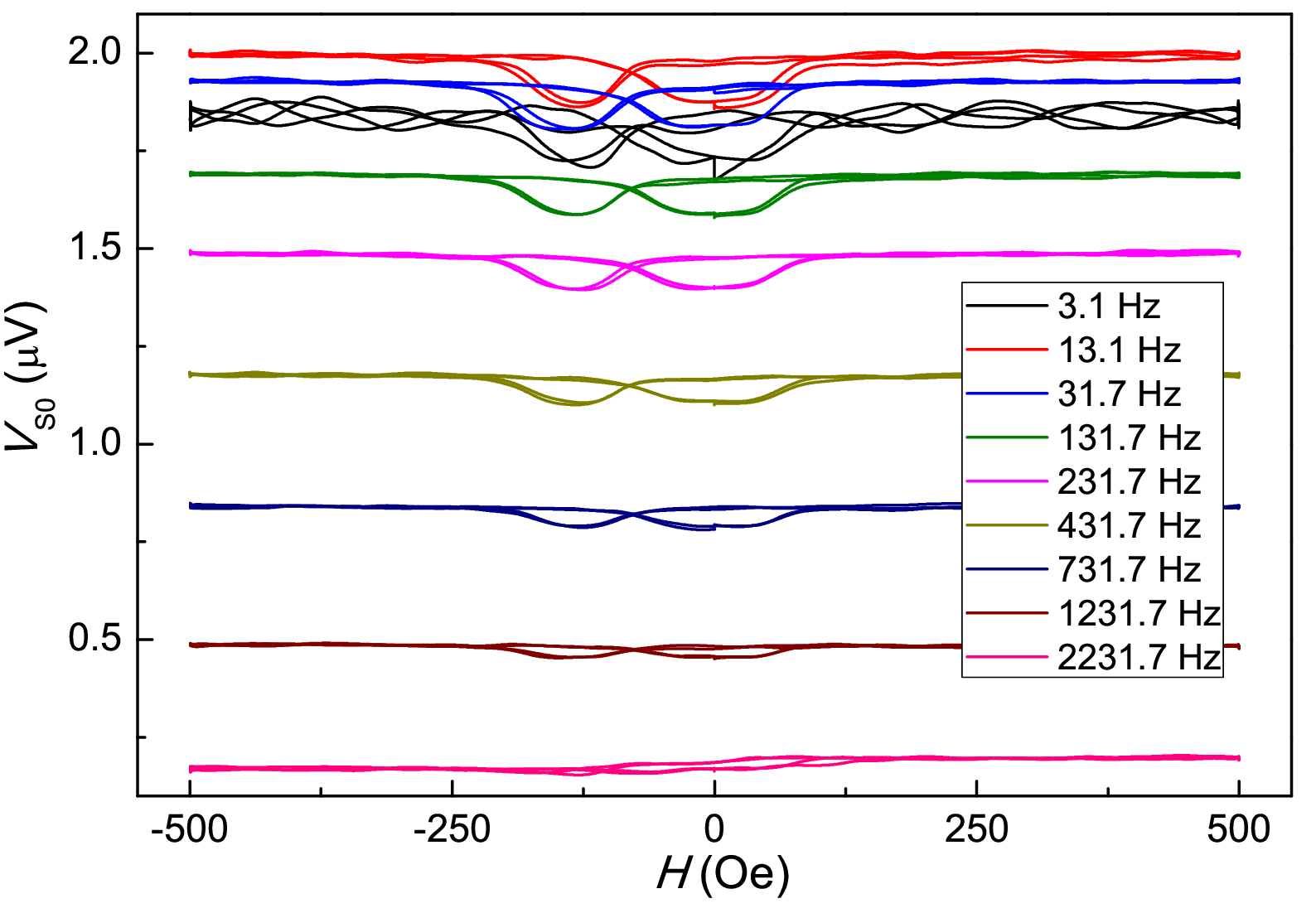}%
\caption{\label{f-S1} (Color online) Field-dependence of Seebeck voltages measured in different frequencies.}
\end{figure}
\begin{figure}
\includegraphics[width=0.45\textwidth]{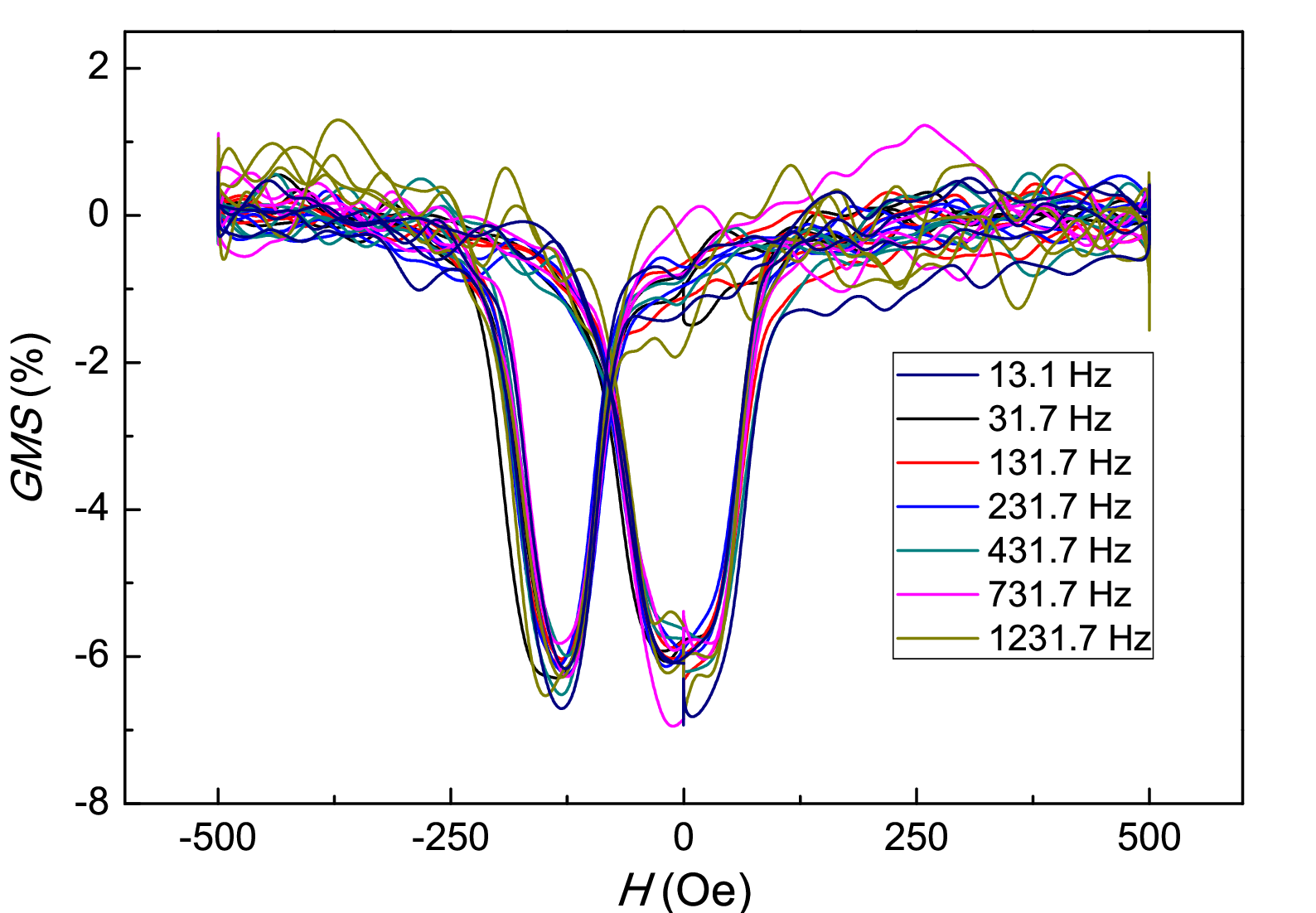}%
\caption{\label{f-S2} (Color online) Field-dependence of GMS ratios measured in different frequencies.}
\end{figure}
\begin{figure}
\includegraphics[width=0.45\textwidth]{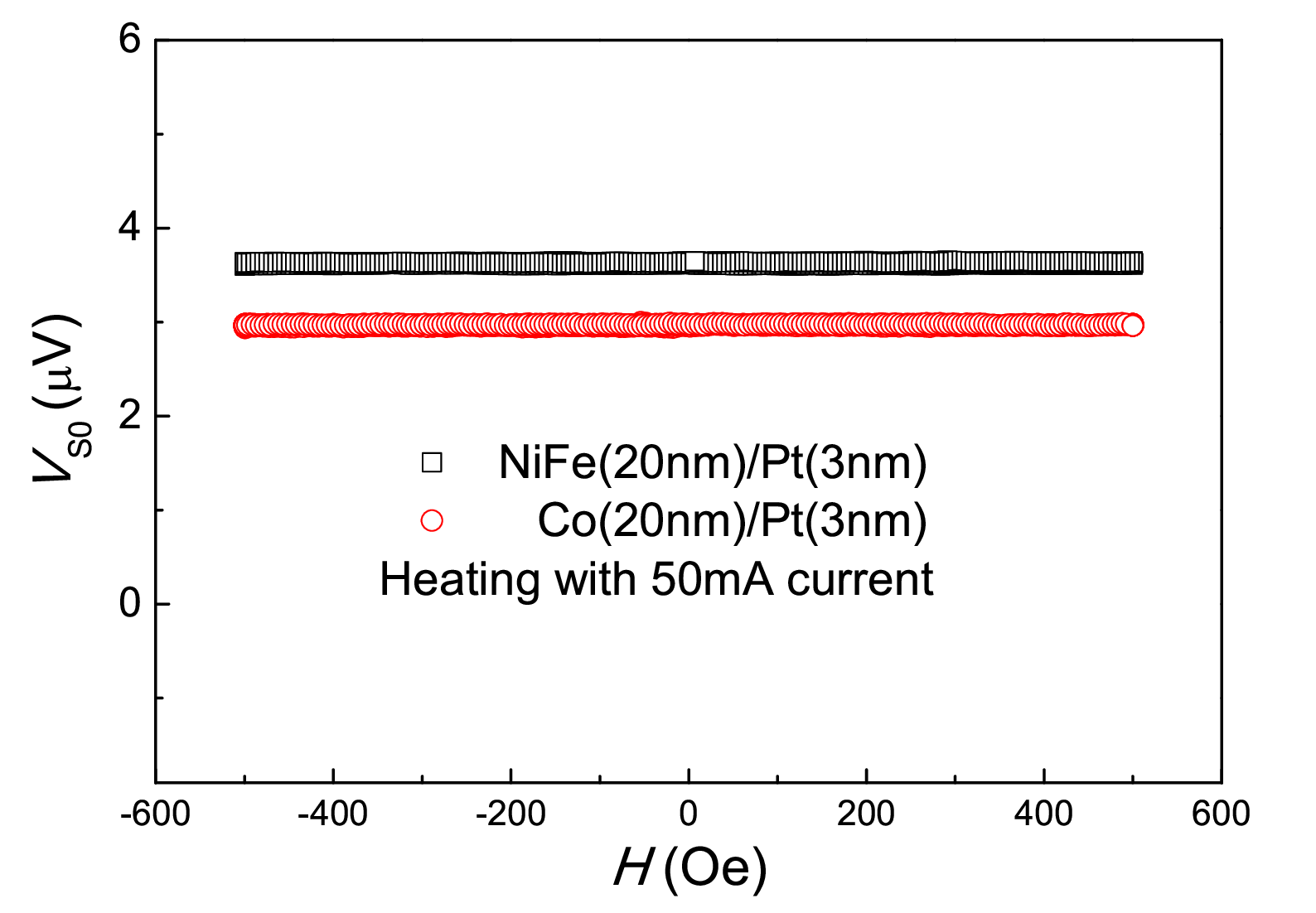}%
\caption{\label{f-S3} (Color online) Field dependence of Seebeck coefficients of thick NiFe and Co films.}
\end{figure}

\textbf{C. $T$- and $H$-dependence of GMR and GMS ratios of Co/Cu/Co spin valves}

Figure S4(a) and Figure S5(a) shows that GMR ratios increased linearly with decreasing temperatures and the GMR ratio increased from 5.5\% at 300 K to 9\% at 50 K while absolute value of GMS ratio decreased with decreasing temperatures, from 6\% at 300 K to 0\% at 50 K as shown in Figure S4(b) and Figure S5(b). Same as the results at room temperature, $GMR>0$ while $GMS<0$ at low temperatures as well. d$GMR$/d$T$=-1.5\%/100 K, as occurred in other GMR spin valves \cite {Torres}. Regarding saturated GMS ratios at high temperatures, we think it was related with energy-dependence of GMR ratios as following discussion. According to Mott relation \cite {Cutler}, $S=S_{0}(k\mathrm{_{B}}T)$(dln$\sigma$/d$E$)$_{E\mathrm{_{F}}}$, $S_{0}$=$\pi^{2}k\mathrm{_{B}}$/3$e$=283 $\mathrm{\mu V/K}$, where $S$, $k\mathrm{_{B}}$, $T$, $\sigma$, $E$ and $E\mathrm{_{F}}$ is Seebeck coefficient, Boltzmann constant, Kelvin temperature, conductivity, carrier energy and Fermi energy, respectively. $GMS\equiv (S\mathrm{_{AP}}-S\mathrm{_{P}})/S\mathrm{_{P}}$.

-$GMS\cdot S\mathrm{_{P}}=-GMS\cdot S_{0}(k\mathrm{_{B}}T)$(dln$\sigma\mathrm{_{P}}$/d$E$)$_{E\mathrm{_{F}}}$

=$S\mathrm{_{P}}-S\mathrm{_{AP}}$=$S_{0}(k\mathrm{_{B}}T)$(dln$\sigma\mathrm{_{P}}$/d$E$)-$S_{0}(k\mathrm{_{B}}T)$(dln$\sigma\mathrm{_{AP}}$/d$E$)

=$S_{0}(k\mathrm{_{B}}T)$dln($\sigma\mathrm{_{P}}/\sigma\mathrm{_{AP}})$/d$E$=$S_{0}(k\mathrm{_{B}}T)$dln(1+$GMR$)/d$E$.

In our case, $GMR\ll1$, therefore, d$GMR$/d$E\approx-GMS\times S\mathrm{_{P}}/(k\mathrm{_{B}}TS_{0})$, which indicated that GMS ratio had close relation with energy derivative of GMR ratio. $S\mathrm{_{P}}/T$ in our experiment was nearly constant, about -0.015 V/$\mathrm{K}^{2}$, during 150 K-300 K. Thus d$GMR$/d$E$=-3.7\%/eV, higher energy of carriers resulting in lower GMR ratios, as reported in Ref. \cite {Sato}.

\begin{figure}
\includegraphics[width=0.45\textwidth]{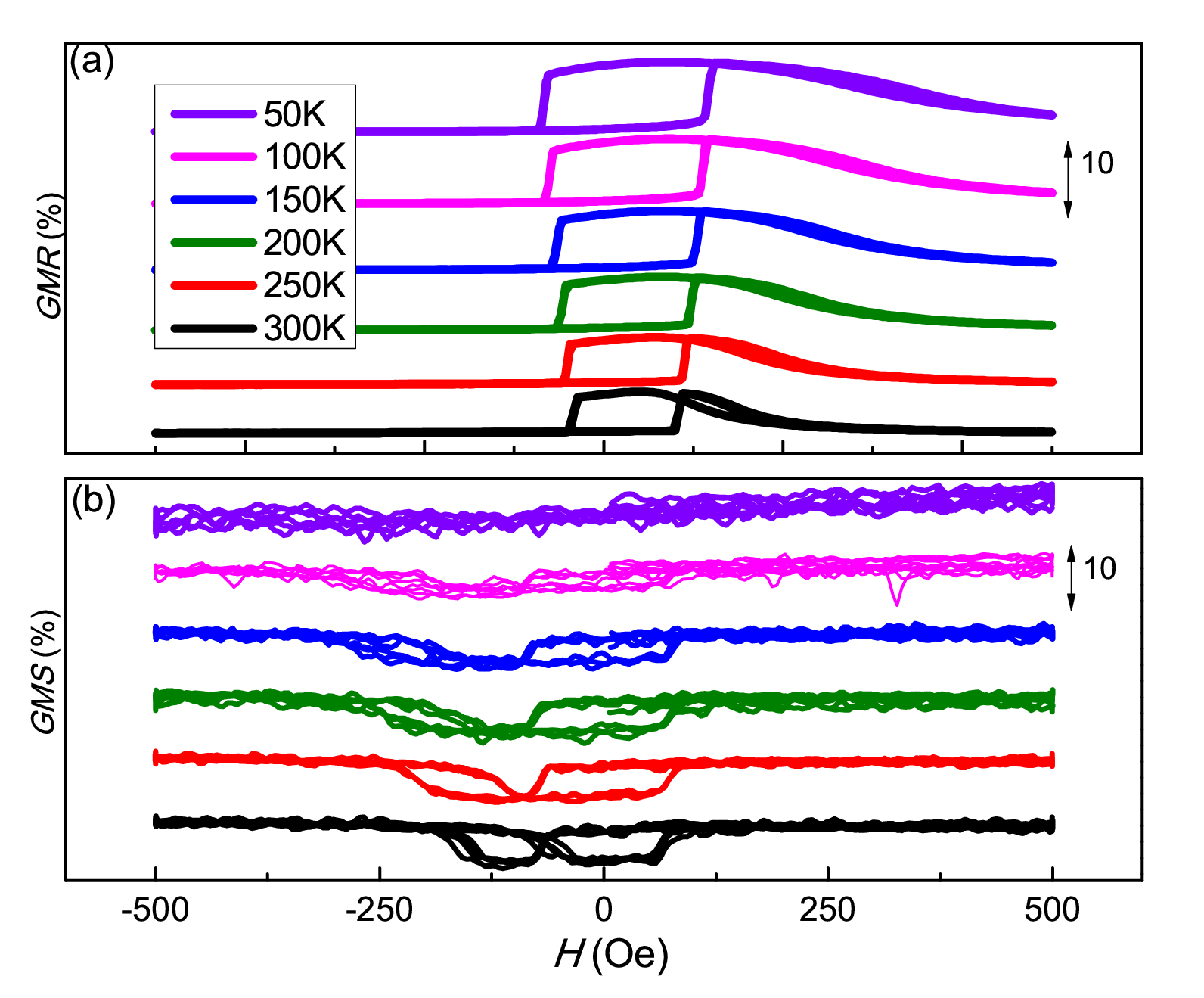}%
\caption{\label{f-S4} (Color online)Field-dependence of GMR and GMS ratios under different temperatures}
\end{figure}

\begin{figure}
\includegraphics[width=0.45\textwidth]{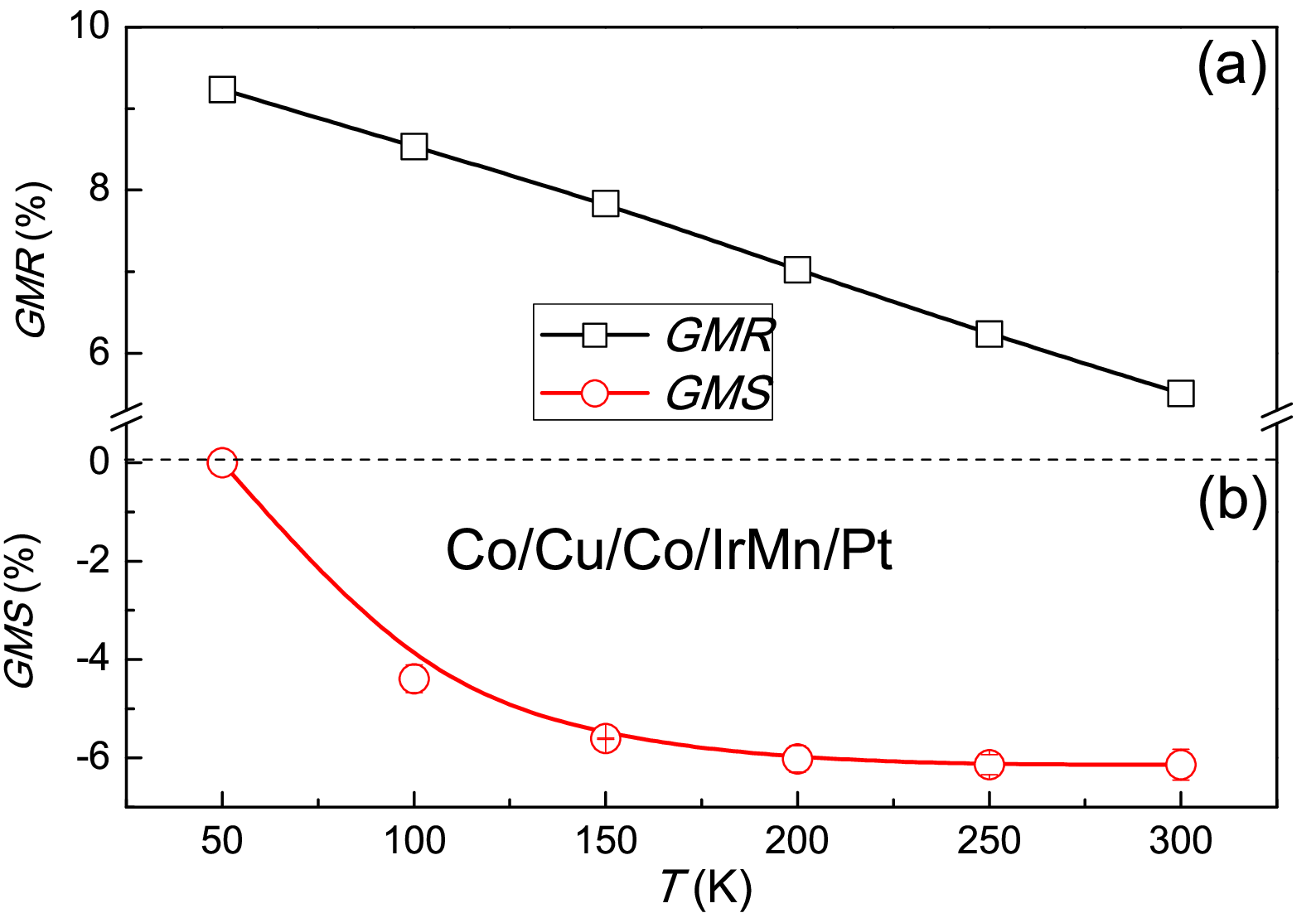}%
\caption{\label{f-S5} (Color online)Temperature dependence of maximum GMR and GMS of Co/Cu/Co/IrMn/Pt spin valves.}
\end{figure}